\documentstyle[12pt,aps]{revtex}
\tightenlines

\begin{document}

\begin{center}
{\LARGE Edge helicons and repulsion of fundamental edge magnetoplasmons in
the quantum Hall regime}

O. G. Balev$^{1,2}$, P. Vasilopoulos$^{3}$, and Nelson Studart$^{1}$
\end{center}

$^{1}$Departmento de Fisica, Universidade Federal de S\~{a}o Carlos,
13565-905 S\~{a}o Carlos,

S\~{a}o Paulo, Brazil

$^{2}$Institute of Physics of Semiconductors, National Academy of Sciences,
45 Prospekt Nauky, Kiev 252650, Ukraine

$^{3}$Concordia University, Department of Physics, 1455 de Maisonneuve Blvd
O, Montr\'{e}al, Qu\'{e}bec, Canada, H3G 1M8

\begin{center}
{\large Abstract}
\end{center}

A {\it quasi-microscopic} treatment of edge magnetoplasmons (EMP) is
presented for very low temperatures and confining potentials smooth on the
scale of the magnetic length $\ell _{0}$ but sufficiently steep at the edges
such that Landau level (LL) flattening can be discarded. The profile of the
unperturbed electron density is sharp and the dissipation taken into account
comes only from electron intra-edge and intra-LL transitions due to
scattering by acoustic phonons. For wide channels and filling factors $\nu =1
$ and $2$, there exist independent EMP modes spatially symmetric and
antisymmetric with respect to the edge. Some of these modes, named edge
helicons, can propagate nearly undamped even when the dissipation is strong.
Their density profile changes qualitatively during propagation and is given
by a rotation of a complex vector function. For $\nu >2,$ the Coulomb
coupling between the LLs leads to a repulsion of the uncoupled fundamental
LL modes: the new modes have very different group velocities and are nearly
undamped. The theory accounts well for the experimentally observed plateau
structure of the delay times as well as for the EMP's period and decay
rates.\ \newline
PACS\ \ 73.20.Dx, 73.40.Hm

\section{INTRODUCTION}

Previous theoretical studies of edge magnetoplasmons (EMP), the
low-frequency collective excitations which propagate along the edges of a
two-dimensional electron gas (2DEG) subject to a normal magnetic field $B$,
point out some important characteristics of EMP, e.g., the gapless
excitation spectrum$^{\cite{1}}$ and the acoustic EMP.$^{\cite{2}}$ However,
the authors of Refs. $\cite{1}$ and $\cite{2}$ have assumed density profiles
which are infinitely sharp or smooth respectively and independent of the
filling factor $\nu =n_{0}h/|e|B$, where $n_{0}$ is the electron density in
the bulk of 2DEG. As a consequence, they do not reflect the Landau-level
(LL) structure, as one can see in Fig. 1, in which we compare our calculated
density profile for the cases of one and two occupied LLs, by assuming a
smooth parabolic confining potential at the edge, with the results of Refs. $%
\cite{1}$ and $\cite{2}$. This inadequacy is clearly manifested in the
observed$^{\cite{3}}$ plateau structure of the transit times reflecting that
of the quantum Hall effect (QHE) plateaus and not accounted for in Ref.$\cite
{2}$. In addition, for a spatially homogeneous dissipation within the
channel, the EMP damping was found quantized and independent of temperature$%
^{\cite{1}}$ or it was treated phenomenologically.$^{\cite{2}}$ The
calculated damping rates were strongly overestimated$^{\cite{3}}$. Other
drawbacks and the limited validity of the treatment of Ref.$\cite{1}$ in the
QHE regime were pointed out in Refs.$\cite{4}$-$\cite{6}$.

In this paper, we present a theory of EMP for integer $\nu $ in which the LL
structure is taken into account. In doing so, we have assumed that the
confining potential is sufficiently steep at the edges such that the
dissipation is significant$^{\cite{7},\cite{8}}$ only within a distance from
the edge of the order of the magnetic length $\ell _{0}=\sqrt{\hbar /|e|B}$
and that LL flattening$^{\cite{9}}$ can be neglected.$^{\cite{10},\cite{11}}$
We show that, for $\nu >2$, the Coulomb coupling, between the charge
excitations at the edges of different occupied LLs leads to a repulsion of
the uncoupled fundamental LL modes. The new modes, which are nearly
undamped, have very different group velocities.

For the 2D system with a vertical conductivity drop at the boundaries, it
has been shown$^{\cite{1}}$ that the dissipation determines significantly
the dispersion relation and the spatial structure of the EMP even in the
regime of the QHE. The properties of the EMP have been expressed in terms of
the components of the magnetoconductivity tensor of an infinite 2D system.
Moreover, due to the very low frequency $\omega $ of the EMP, the dispersion
relation could be written in terms of the static magnetoconductivity tensor.
However, in Refs.$\cite{7}$ and $\cite{8}$ we have shown that, for a
sufficiently smooth lateral confinement and in the QHE regime, the
dissipation appears dominantly due to intralevel-intraedge transitions of
electrons interacting with acoustic phonons and occurs mainly near the
edges. For instance, in GaAs-based samples, piezoelectrical (PA) phonons are
typically involved. In the linear response regime this is the main cause for
dissipation in channels of width $W\alt 100\ \mu $m and temperatures $T\alt 1
$K if the group velocity of the edge states $v_{g}$ is larger than the speed
of sound $s$. As in the case of dissipation in the bulk, it is exponentially
suppressed as the temperature goes to zero. Under this circumstance, and the
fact that the dissipation in the Volkov-Mikhailov model$^{\cite{1}}$ is
homogeneous over the channel width, the previous results of Balev and
Vasilopoulos$^{\cite{12},\cite{13}}$ are understood as a demonstration that
the EMP properties reported in Ref.$\cite{1}$ can be strongly modified by
dissipative processes localized near the channel edges. Here we assume very
low temperatures which implies that the condition $k_{B}T\ll \hbar
v_{g}/\ell _{0}$ is fulfilled. A brief account of some results of the
present investigation has appeared in Ref.$\cite{13}$.

The organization of the paper is as follows. In Sec. II, by starting with
the expressions for the inhomogeneous current densities and conductivities,
we derive the integral equation for EMPs and present the general method for
solving it. In Sec. III we calculate the dispersion relations and the
spatial structure of the new edge waves at very low temperatures. Finally,
in Sec. IV we compare our theory with experiment and present our concluding
remarks.

\section{BASIC RELATIONS}

\subsection{Inhomogeneous current density in the quasi-static regime}

We consider a 2DEG confined to a strip in the $x-y$ plane with a width $W$
in the $y$ axis and length $L$ along a channel in the $x$ direction, under a
strong magnetic field $B$ parallel to the $z$ axis. For simplicity, we
consider the confining potential as parabolic at the edges such that $%
V_{y}=0 $, for $y_{l}<y<y_{r}$, $V_{y}=m^{*}\Omega ^{2}(y-y_{r})^{2}/2$ for $%
y>y_{r}>0$, and $V_{y}=m^{*}\Omega ^{2}(y-y_{l})^{2}/2$ for $y<y_{l}<0$,
where $y_{r,l}$ delimited the right and left edges of the flat part of $V_{y}
$. We assume the condition $|k_{x}|W \gg 1$, so it is possible to consider
an EMP along the right edge of the channel of the form $A(\omega
,k_{x},y)\exp [-i(\omega t-k_{x}x)]$ totally independent of the left edge.
We consider only the linear response regime. For definiteness, we take the
background dielectric constant $\epsilon $ to be spatially homogeneous. In
the QHE regime and $\nu $ even, we neglect the Zeeman spin splitting. As for
the case $\nu =1$, we assume that the spin splitting, caused by many-body
effects, is strong enough to neglect the contribution related to the upper
spin-split LL. We also assume a smooth lateral confinement on the scale of
the magnetic length $\ell_{0}=(\hbar /m^{*}\omega _{c})^{1/2}$ such that $%
\Omega \ll \omega _{c}$, where $\omega _{c}=|e|B/m^{*}$ is the cyclotron
frequency.

The EMP is practically quasi-static and its wavelength $\lambda $ is much
larger than $\ell _{0}$. As in Refs. $\cite{12}$ and $\cite{13}$, we write
the components of the current density in the form

\begin{eqnarray}
j_{y}(y) &=&\sigma _{yy}(y)E_{y}(y)+\sigma _{yx}^{0}(y)E_{x}(y),  \label{2}
\\
&&  \nonumber
\end{eqnarray}

\begin{equation}
j_{x}(y)=\sigma _{xx}(y)E_{x}(y)-\sigma
_{yx}^{0}(y)E_{y}(y)+\sum_{j}v_{gj}\rho _{j}(\omega ,k_{x},y),  \label{3}
\end{equation}
where $\sigma _{\mu \gamma}(y)$ and $E_{\gamma}(y)$ are the components of
the conductivity tensor and the electric field respectively. Here we have
suppressed the exponential factor $\exp [-i(\omega t-k_{x}x)]$ common to all
terms in Eqs. (\ref{2}) and (\ref{3}). It is understood that $E_{\gamma}(y)$
depends on $\omega $ and $k_{x}$. The term $v_{gj}\rho _{j}(\omega ,k_{x},y)$
represents an advection contribution caused by a charge distortion $\rho
_{j}(\omega,k_{x},y)$ localized near the edge $y_{rj}$ of the $j$-th LL. In
Eqs. (\ref{2}) and (\ref{3}) the contributions to $j_{\mu }(y)$, which are
proportional to $E_{\gamma}(y)$, are microscopically obtained when the
electric field is smooth on the scale of $\ell _{0}$. Even though this is
not well justified for both components proportional to $E_{x}(y)$ and $%
E_{y}(y)$, we approximate these contributions by those obtained when $%
E_{x}(y)$ and $E_{y}(y)$ are smooth on the scale of $\ell _{0}$. This
approximation is equivalent to neglecting nonlocal contributions to $j_{\mu}
\propto \int dy^{\prime } \sigma _{\mu \gamma }(y,y^{\prime
})E_{\gamma}(y^{\prime })$. For weak dissipation, it can be justified within
a treatment based on the random-phase-approximation (RPA)$^{\cite{14}}$ that
includes nonlocal effects and edge-state screening, e.g., for the
fundamental EMPs at $\nu =2,4,6$. The Hall conductivity is$^{\cite{8}}$

\begin{equation}
\sigma _{yx}^{0}(y)=\frac{e^{2}}{2\pi \hbar }\sum_{n=0}\int_{-\infty
}^{\infty }dy_{0\alpha }f_{\alpha }\Psi _{n}^{2}(y-y_{0\alpha }),  \label{4}
\end{equation}
where $\alpha \equiv \{n,k_{x\alpha }\}$,\ $y_{0\alpha }=\ell
_{0}^{2}k_{x\alpha }$, $\Psi _{n}(y)$ is the harmonic oscillator function,
and $f_{\alpha }\equiv f_{n}(k_{x\alpha })=1/[1+\exp (\varepsilon _{\alpha
}-\varepsilon _{F})/k_{B}T]$ is the Fermi-Dirac function. $\varepsilon _{F}$
is the Fermi level measured from the bottom of the lowest electric subband;
for even $\nu $ the right-hand side of Eq. (\ref{4}) should be multiplied by
2, the spin degeneracy factor. We point out that, for $\nu =1$ and $T=0$ and
near the right edge, we obtain $\sigma _{yx}^{0}(y)=(e^{2}/4\pi \hbar
)[1+\Phi (y_{re}-y)]$, where $\Phi (x)$ is the probability integral, $%
y_{re}=\ell _{0}^{2}k_{re}$, and $f_{0}(k_{re})=1/2$. Notice then that $%
\sigma _{yx}^{0}(y)$, near the edge, decreases on the scale of $\ell _{0}$
and behaves like the density profile depicted by the dashed curve of Fig. 1.
Considering only the right edge and the flat part of the confining potential
for $y_{l}\leq y_{0\alpha }\leq y_{r}$, we obtain the energy levels $%
\varepsilon _{\alpha }=\hbar \omega _{c}(n+1/2)$ and for $y_{0\alpha } \geq
y_{r}$, the energy spectrum can be written as

\begin{equation}
\varepsilon _{\alpha }\equiv \varepsilon _{n}(k_{x\alpha })=\hbar \omega
_{c}(n+1/2)+m^{*}\Omega ^{2}(y_{0\alpha }-y_{r})^{2}/2.  \label{5}
\end{equation}
This results implies that $\varepsilon _{\bar{n}}(k_{x\alpha })$, as a
function of $y_{0\alpha }$, is smooth on the scale of $\sqrt{(2\bar{n}+1)}%
\ell _{0}$, where $\bar{n}$ is the principal quantum number of the highest
occupied LL. The energy spectrum (\ref{5}) of the $n$-th LL allows us to
evaluate the group velocity of the edge states as $v_{gn}=\partial
\varepsilon _{n}(k_{r}+k_{e}^{(n)})/\hbar \partial k_{x}=\hbar \Omega
^{2}k_{e}^{(n)}/m^{*}\omega _{c}^{2}$ with the characteristic wave vector $%
k_{e}^{(n)}=(\omega _{c}/\hbar \Omega )\sqrt{2m^{*}\Delta _{Fn}}$, where $%
\Delta _{Fn}=\varepsilon _{F}-(n+1/2)\hbar \omega _{c}$. The edge of the $n$%
-th LL is denoted by $y_{rn}=y_{r}+\ell _{0}^{2}k_{e}^{(n)}=\ell
_{0}^{2}k_{rn}$, where $k_{rn}=k_{r}+k_{e}^{(n)}$, $k_{r}=y_{r}/\ell
_{0}^{2} $ and $W=2y_{r0}$. We can also write $v_{gn}=E_{en}/B$, where $%
E_{en}=\Omega \sqrt{2m^{*}\Delta _{Fn}}/|e|$ is the electric field
associated with the confining potential $V_{y}$ at $y_{rn}$.

We consider only the electron-phonon interaction and neglect that of
electrons with impurities, since the former is the most essential for the
assumed conditions$^{\cite{7}}$. Following Refs. $\cite{12}$ and $\cite{13}$%
, we approximate $\sigma _{xx}(y)$ by $\sigma _{yy}(y)=\sum_{n=0}^{\bar{n}%
}\sigma _{yy}^{(n)}(y)$. Furthermore, at very low $T$ where $\hbar v_{gn}
\gg \ell _{0}k_{B}T$ with $v_{gn}>s$ is satisfied, the Eq. ($16$) of Ref. $%
\cite{8}$ gives $\sigma _{yy}^{(n)}(y)=\tilde{\sigma}_{yy}^{(n)}\Psi
_{n}^{2}(\bar{y}_{n})$, $\bar{y}_{n}=y-y_{rn}$. For PA phonons and $\nu =2,4$%
, we have $\tilde{\sigma}_{yy}^{(n)}=3e^{2}\ell _{0}^{4}c^{\prime
}k_{B}^{3}T^{3}/\pi ^{2}\hbar ^{6}v_{gn}^{4}s$ where $c^{^{\prime }}$ is the
electron-phonon coupling constant. Note then that $\sigma _{yy}(y)$ is
exponentially localized within a distance $\alt \ell _{0}$ from the LL edge $%
y_{rn}$. Notice also that for $v_{gn}/s\gg 1$ such that $1\gg k_{B}T\ell
_{0}/\hbar v_{gn}>s/\sqrt{2}v_{gn}$ we have $\tilde{\sigma}_{yy}^{(n)}=\sqrt{%
2}e^{2}\ell _{0}^{3}c^{\prime }k_{B}^{2}T^{2}/\pi ^{5/2}\hbar ^{5}v_{gn}^{4}$%
. Furthermore, for all cases it is assumed that the strong magnetic field
condition $\sigma _{yy}(y)/|\sigma _{yx}^{0}(y)|\ll 1$ is fulfilled.

\subsection{Integral equation for EMPs with dissipation at the edges}

Using Eqs. (\ref{2})-(\ref{4}), the Poisson's equation, and the linearized
continuity equation, we obtain the following integral equation for $\rho
(\omega ,k_{x},y)$

\begin{eqnarray}
-i\sum_{n} &&(\omega -k_{x}v_{gn})\rho _{n}(\omega ,k_{x},y)+\frac{2}{%
\epsilon }\Big[ k_{x}^{2}\sigma _{xx}(y)  \nonumber \\
* &&-ik_{x}\frac{d}{dy}[\sigma _{yx}^{0}(y)]-\sigma _{yy}(y)\frac{d^{2}}{%
dy^{2}}-\frac{d}{dy}[\sigma _{yy}(y)]\frac{d}{dy}\Big]  \nonumber \\
* &&\times \int_{-\infty }^{\infty }dy^{\prime }K_{0}(|k_{x}||y-y^{\prime
}|)\rho (\omega ,k_{x},y^{\prime })=0,  \label{6}
\end{eqnarray}
where $K_{0}(x)$ is the modified Bessel function. For the dissipationless
classical 2D electron liquid Eq. (\ref{6}) becomes identical with Eq. (4) of
Ref. $\cite{2}$. Furthermore, if the conductivity components are independent
of $y$, for $|y|<W/2$, Eq. (\ref{6}) assumes the form of Eq. (15) of Ref. $%
\cite{1}$. In order to solve Eq. (\ref{6}), we remark that, for $\hbar
v_{gn}\gg \ell _{0}k_{B}T$, we have $d[\sigma _{yx}^{0}(y)]/dy\propto
\sum_{n=0}^{\bar{n}}\Psi _{n}^{2}(\bar{y}_{n})$, whose spatial behavior is
similar to that of $\sigma _{yy}(y)$. Then Eq. (\ref{6}) shows that $\rho
_{n}(\omega ,k_{x},y)$ will be concentrated within a region of extent $\sim
\ell _{0}$ around the edge of the $n$-th LL. Further, the integral can be
evaluated under the assumption $k_{x}\ell _{0}\ll 1$ and using the
approximation $K_{0}(|x|)\approx \ln (2/|x|)-\gamma $, where $\gamma $ is
the Euler constant. Assuming $\Delta y_{m-1,m}=y_{rm-1}-y_{rm}\gg \ell _{0}$%
, (see Fig. 1 for $\nu =4$), we can neglect the exponentially small overlap
between $\rho _{m-1}(\omega ,k_{x},y)$ and $\rho _{m}(\omega ,k_{x},y)$, for 
$m\leq \bar{n}$. It is then natural to attempt the exact solution in the form

\begin{equation}
\rho (\omega ,k_{x},y)=\sum_{n=0}^{\bar{n}}\rho _{n}(\omega
,k_{x},y)=\sum_{n=0}^{\bar{n}}\Psi _{n}^{2}(\bar{y}_{n})\sum_{l=0}^{\infty
}\rho _{n}^{(l)}(\omega ,k_{x})H_{l}(\bar{y}_{n}/\ell _{0}),  \label{7}
\end{equation}
by using the Hermite polynomials $H_{l}(x)$ as a expansion basis. We call
the terms $l=0,l=1,l=2$, and so on, the monopole, dipole, quadrupole terms
in the expansion of $\rho _{n}(\omega ,k_{x},y)$ relative to $y=y_{rn}$.

\subsection{EMPs for $\nu =2$ and $\nu =4$}

Below we first give the general formulas in the case $\nu =4$ ($\bar{n}=1$).
>From them, we show how general formulas for $\nu =2$ ($\bar{n}=0$) follow
up. For $\nu =4$ we multiply Eq. (\ref{6}) by $H_{m}(\bar{y}_{0}/\ell _{0})$
and integrate over $y$, from $y_{r0}-\Delta y_{01}/2$ to $y_{r0}+\Delta
y_{01}/2$. Analogous integration, from $y_{r1}-\Delta y_{01}/2$ to $%
y_{r1}+\Delta y_{01}/2$, is repeated with $H_{m_{1}}(\bar{y}_{1}/\ell _{0})$%
. With the abbreviations $\rho _{0}^{(m)}(\omega ,k_{x})\equiv \rho_{0}^{(m)}
$, $a_{mk}(k_{x})\equiv a_{mk}$, etc., we obtain, the following coupled
systems of equations

\begin{eqnarray}
(\omega-k_{x}v_{g0})\rho_{0}^{(m)}&& -(S_{0}+mS_{0}^{`})
\sum_{n=0}^{\infty}c_{mn}a_{mn}\rho_{0}^{(n)}  \nonumber \\
* &&-(S_{0}+mS_{0}^{`}) \sum_{l=0}^{\infty}c_{ml}b_{ml}\rho_{1}^{(l)} =0,
\label{8}
\end{eqnarray}

\begin{eqnarray}
&&(\omega-k_{x}v_{g1})\Big[A_{m_{1}}\rho_{1}^{(m_{1})}+B_{m_{1}}
\rho_{1}^{(m_{1}+2)}+\rho_{1}^{(m_{1}-2)}/2\Big]  \nonumber \\
* &&-(S_{1}+m_{1}S_{1}^{`}) \Big[\sum_{n=0}^{\infty}c_{m_{1}n} \
b_{nm_{1}}\rho_{0}^{(n)}
+\sum_{j=0}^{\infty}c_{m_{1}j}d_{m_{1}j}\rho_{1}^{(j)}\Big]  \nonumber \\
* && +2\sqrt{m_{1}}S_{1}^{`}\Big[ \sum_{n=0}^{\infty}c_{m_{1}n} \ \tilde{b}%
_{n,m_{1}}\rho_{0}^{(n)} +\sum_{j=0}^{\infty}c_{m_{1}j}\tilde{d}_{m_{1},j}
\rho_{1}^{(j)}\Big]=0,  \label{9}
\end{eqnarray}
where

\begin{eqnarray}
a_{mn}=&&a_{nm} =\int_{-\infty}^{\infty} dx\ \Psi_{m}(x)\Psi_{0}(x)
\int_{-\infty}^{\infty} dx^{\prime}  \nonumber \\
* &&\times K_{0}(|k_{x}||x-x^{\prime}|)\ \Psi_{n}(x^{\prime})\Psi_{0}
(x^{\prime}) ,  \label{10}
\end{eqnarray}

\begin{eqnarray}
b_{mn} &=&\int_{-\infty }^{\infty }dx\ \Psi _{m}(x)\Psi _{0}(x)\int_{-\infty
}^{\infty }dx^{\prime }  \nonumber \\
* &&\times 2(x^{\prime }/\ell _{0})^{2}K_{0}(|k_{x}||x-x^{\prime }+\Delta
y_{01}|)\ \Psi _{n}(x^{\prime })\Psi _{0}(x^{\prime }).  \label{11}
\end{eqnarray}
Other coefficients are given as

\begin{eqnarray}
\tilde{b}_{mn}=&&\int_{-\infty}^{\infty} dx\ \Psi_{m}(x)\Psi_{0}(x)
\int_{-\infty}^{\infty} dx^{\prime}  \nonumber \\
* &&\times K_{0}(|k_{x}||x-x^{\prime}+\Delta y_{01}|)
\Psi_{n-1}(x^{\prime})\Psi_{1} (x^{\prime}) ,  \label{12}
\end{eqnarray}

\begin{eqnarray}
d_{mn}=&&(2/\ell_{0}^{2}) \int_{-\infty}^{\infty} dx\ \Psi_{m}(x) \; x \;
\Psi_{1}(x) \int_{-\infty}^{\infty} dx^{\prime}  \nonumber \\
* &&\times K_{0}(|k_{x}||x-x^{\prime}|)\ \Psi_{n}(x^{\prime}) \;
x^{\prime}\; \Psi_{1} (x^{\prime}) ,  \label{13}
\end{eqnarray}

\begin{eqnarray}
\tilde{d}_{mn} &=&(\sqrt{2}/\ell _{0})\int_{-\infty }^{\infty }dx\ \Psi
_{m-1}(x)\Psi _{1}(x)\int_{-\infty }^{\infty }dx^{\prime }  \nonumber \\
* &&\times K_{0}(|k_{x}||x-x^{\prime }|)\ \Psi _{n}(x^{\prime })\;x^{\prime
}\;\Psi _{1}(x^{\prime }).  \label{14}
\end{eqnarray}
In addition, $c_{mn}=\sqrt{2^{n}n!/2^{m}m!}$, $A_{m_{1}}=(2m_{1}+1)$, $%
B_{m_{1}}=(m_{1}+2)(2m_{1}+2)$, $S_{j}=2(k_{x}\tilde{\sigma}%
_{yx}^{0}-ik_{x}^{2}\tilde{\sigma}_{xx}^{(j)})/\epsilon $, $S_{j}^{`}=-4i%
\tilde{\sigma}_{yy}^{(j)}/\epsilon \ell _{0}^{2}$, and $\tilde{\sigma}%
_{yx}^{0}=e^{2}/\pi \hbar $. Notice that here in the interior part of the
channel $\sigma _{yx}^{0}(y)=\sigma _{yx}^{0}=2\tilde{\sigma}%
_{yx}^{0}=2e^{2}/\pi \hbar $.

Furthermore, for $\nu =2$ ($\bar{n}=0$) the third term in Eqs. (\ref{8}) and
(\ref{9}) is absent and we obtain explicitly

\begin{equation}
(\omega -k_{x}v_{g0})\rho _{0}^{(m)}-(S_{0}+mS_{0}^{`})\sum_{n=0}^{\infty
}c_{mn}a_{mn}\rho _{0}^{(n)}=0.  \label{15}
\end{equation}
In this case we obtain $\sigma _{yx}^{0}(y)=\sigma _{yx}^{0}=\tilde{\sigma}%
_{yx}^{0}=e^{2}/\pi \hbar $ in the inner part of the channel.

\section{Edge waves}

\subsection{Edge modes for $\nu =2(1)$}

In what follows we assume $\nu =2$ but the results obtained can be easily
extended to the case of $\nu =1$, when only the lowest spin-split LL is
occupied. Equations (\ref{7}) and (\ref{15}) show that there exist
independent modes, spatially {\it symmetric}, $\rho ^{s}(\omega ,k_{x},y)$,
or {\it antisymmetric}, $\rho ^{a}(\omega ,k_{x},y)$, with respect to $%
y=y_{r0}$ (see also Ref. $\cite{12})$. They correspond to $l$ even or odd,
respectively, in Eq. (\ref{7}). Notice that in this case $\bar{n}=0$ and all
edge modes belong to the $n=0$ LL. However, within our quasi-microscopic
approach they typically include mixing of the $n=0$ LL with some higher
empty LLs.

We first consider the lowest antisymmetric mode for {\it weak dissipation}$^{%
\cite{12}}$ $\eta =\tilde{\sigma}_{yy}^{(0)}/(\ell _{0}^{2}
\sigma_{yx}^{0}|k_{x}|) \ll 1/4$. As we neglect its coupling with higher
antisymmetric modes, it becomes purely dipole and corresponds to $l=1$ and $%
n=0$ in Eq. (\ref{7}). To take into account the effect of inter-mode
coupling on this pure dipole ($l=1$) mode we will neglect damping. Then, as
shown in Ref.$\cite{12}$, the pure dipole mode has a dimensionless velocity $%
v_{dip}=(\omega /k_{x}-v_{go})/ (2\sigma _{yx}^{0}/\epsilon )$ given by $%
v_{dip}=a_{11} \approx 0.4996$. Furthermore, if only the interaction of this
mode with the octupole mode is considered, the velocity of the resulting
renormalized dipole mode becomes $v_{dip}=(a_{11}+a_{33}+\sqrt{%
(a_{11}-a_{33})^{2}+4a_{13}^{2}})/2\approx 0.5963$. Here only the $l=1$ and $%
l=3$ terms in Eq. (\ref{7}) are involved. It is seen that due to the
interaction with the $l=3$ mode, $v_{dip}$ becomes $\approx 20\%$ higher. To
take into account the interaction of the dipole mode with the $l=3$ and $l=5$
modes, we retain the terms $l=1,3$ and $l=5$ in Eq. (\ref{7}). From Eq. (\ref
{15}), for $m=1,3$ and $m=5$, we obtain a system of three linear equations
for $\rho _{0}^{(1)}$, $\rho _{0}^{(3)}$, and $\rho _{0}^{(5)}$ that leads
to $v_{dip}\approx 0.6287$. Thus, the interaction with the $l=5$ mode led to
a $v_{dip}$ only $\approx 5\%$ higher than that where only the terms with $%
l=1,3$ were retained. Hence $v_{dip}$, and consequently the dispersion
relation of the dipole mode, exhibits fast convergence when more $l$ terms
are taken into account. Notice that, within the present quasi-microscopic
approach, the dipole mode has a purely acoustic dispersion relation. The
charge-density profile $\delta \rho (y)$ of the dipole mode is shown in Fig.
2 where we plot $\delta \rho (y)\equiv \tilde{\rho}(v_{dip},y)=\sqrt{\pi }
\ell _{0}\rho (\omega _{dip}(k_{x}),k_{x},y)/\rho ^{(1)}(\omega_{dip}
(k_{x}),k_{x})$ as a function of $\bar{y}_{0}/\ell_{0}$. The dashed,
short-dashed, and solid curves are obtained, respectively, with one ($l=1$),
two ($l=1,3$) or three ($l=1,3,5$) terms retained in the expansion (\ref{7}%
). The profile shows a clear convergence already for $l\leq 5$.

For {\it very strong dissipation} $\eta \gg K$, where $K=\ln (1/|k_{x}
\ell_{0}|)+1/2$, we consider only the weakly damped EMP branch that was
termed low-frequency edge helicon (LFEH) in Refs. $\cite{12}$ and $\cite{13}$%
. This mode is symmetric and its real part Re $\omega (k_{x})\equiv $ Re $%
\omega _{EH}^{LF}\approx k_{x}v_{g0}+(2\sigma _{yx}^{0}k_{x}/\epsilon
)(K-1/4)$ is very close to Re $\omega (k_{x})$ of the fundamental EMP of the 
$n=0$ LL, namely, Re$\omega _{EH}^{(0)}\approx k_{x}v_{g0}+(2\sigma
_{yx}^{0}k_{x}/\epsilon )(K+1/4)$. The fundamental EMP of the $n=0$ LL has a
dispersion $\omega _{EH}^{(0)}=k_{x}v_{g0}+S_{0}\ (K+1/4)+S_{0}^{`}/4K$ for
both {\it weak dissipation}, $\eta \ll 1/4$, and {\it strong dissipation}, $%
K\gg \eta \gg 1/4$, [cf. Refs. $\cite{12}$ and $\cite{13}$]. In the latter
case the fundamental EMP was termed$^{\cite{12},\cite{13}}$ {\it %
high-frequency edge helicon} (HFEH). In both regimes of dissipation, the
fundamental EMP of $n=0$ LL has mainly a monopole character$^{\cite{12},\cite
{13}}$: in the first case its profile satisfies $|\rho _{0}^{(2)}/\rho
_{0}^{(0)}| \approx (1/8K) \ll 1$, and in the second, $|\rho _{0}^{(2)}/\rho
_{0}^{(0)}| \approx (\eta /2K) \ll 1$. In either case the fact of
considering only two terms in Eq. (\ref{7}), $l=0$ and $l=2$, is well
justified. The latter is not the case for LFEH as shown in Ref.$\cite{13}$.

To support the statements made above we present in Figs. 3 and 4 the
evolution of the dimensionless charge density profile of the LFEH given by $%
\delta \rho _{r}=\sqrt{\pi }\ell _{0}\ $Re $[\rho (\omega ,k_{x},y)/
\rho_{0}^{(0)}(\omega ,k_{x})]$ and $\delta \rho _{i}=\sqrt{\pi }\ell _{0}\ $%
Im $[\rho (\omega ,k_{x},y)/\rho _{0}^{(0)}(\omega ,k_{x})]$, respectively,
as one increases the number of terms considered in the expansion for $K/\eta
=0.01$. Notice that as $\delta \rho _{r}$ represents the profile for a
particular phase of the wave, then $\delta \rho _{i}$ represents it for a
phase shifted by $\pm \pi /2$. Since $\delta \rho (y)$ is symmetric with
respect to the edge, only the half part of the profile is shown in Figs. 3
and 4. In Fig. 3, curve 1 represents $\delta \rho _{r}$ when only the $l=0$
term is retained in the expansion and curve 2 shows $\delta \rho _{r}\approx 
\sqrt{\pi }\ell _{0}[\Psi _{0}^{2}(\bar{y}_{0})+\sqrt{2}\Psi _{2}(\bar{y}%
_{0})\Psi _{0}(\bar{y}_{0})]$ when two terms $l=0,2$ are retained. As these
curves are essentially different, more $l$ terms should be considered in Eq.
(\ref{7}). Thus, to describe better the profile $\delta \rho _{r}$ we also
plot curves 3, 4, and 5 obtained, respectively, when 3 to 5 even $l$ terms
are retained in Eq. (\ref{7}). This leads to systems of 3, 4, and 5
equations following from Eq. (\ref{15}). For instance, the curve 5 was
obtained by retaining the terms $m=0,2,4,6,$ and $8$. As it is seen in Fig.
3, keeping 4 or 5 terms in the $l$ summation leads already to a rapid
convergent form of $\delta \rho _{r}$ without altering much its oscillatory
character or changing its magnitude. In Fig. 4, similar results are depicted
for the profile $\delta \rho _{i}$ when 2,3, 4, and 5 even $l$ terms are
retained in Eq. (\ref{7}). Notice that the contribution related to the
monopole term, $l=0$, is absent; since the total edge charge $\int dy\delta
\rho _{i}=0$. Therefore $\delta \rho _{i}$ shows essentially a stronger
oscillatory behavior than $\delta \rho _{r}$ and correspondingly a slower
convergence. However, curves 4 and 5 have approximately the same magnitude
in the region where $|\bar{y}_{0}|/\ell _{0}\leq 1.5$. Moreover, for $|\bar{y%
}_{0}|/\ell _{0}\geq 1.5$ these curves exhibit both same spatial behavior
and magnitude.

In Figs. 5 and 6, we plot the same profiles as in Figs. 3 and 4,
respectively, for $K/\eta =0.1$. Curves 1 and 2 in Fig. 5 shows $\delta \rho
_{r}$ with, respectively, only one ( $l=0$) or two ($l=0,2$) terms retained
in Eq. (\ref{7}). As can be seen, keeping 3,4 or 5 terms in the $l$
summation leads already to a clear convergence in the form of $\delta \rho
_{r}$. Figure 6 shows the profile $\delta \rho _{i}$ and the curves are
marked as in Fig. 4. Because the contribution of the monopole term is
absent, $\delta \rho _{i}$ has an essential oscillatory behavior that is
rather similar for curves 4 and 5 as the spatial positions of the extrema
for both curves almost coincide. The dispersion relation of the LFEH
obtained for cases represented by curve 2 in Figs. 3-6 is given by$^{\cite
{12},\cite{13}}$

\begin{equation}
\omega _{EH}^{LF}=k_{x}v_{g0}+\{(2\sigma _{yx}^{0}k_{x}/\epsilon )-[i\tilde{%
\sigma}_{yy}^{(0)}/\eta ^{2}\ell _{0}^{2}\epsilon ]\}(K-1/4).  \label{16}
\end{equation}
When more $l$ terms are taken into account only the imaginary part Im $%
\omega _{EH}^{LF}$ changes significantly. For instance, with 2, 3, 4, or 5
terms retained and $K/\eta =0.01$, the dispersion relations, in Figs. 3 and
4, are given, respectively, by $\Omega _{EH}^{LF}=\omega
_{EH}^{LF}/S_{0}\approx [(K-0.25)-0.005i]$, $\Omega _{EH}^{LF}\approx
[(K-0.50)-0.062i]$, $\Omega _{EH}^{LF}\approx [(K-0.63)-0.117i]$, and $%
\Omega _{EH}^{LF}\approx [(K-0.57)-0.133i]$. It is clearly seen that with 4
and 5 even terms taken into account Im $\omega _{EH}^{LF}$ shows rapid
convergence to its exact value. Despite the {\it very strong dissipation}
the LFEH is {\it very weakly damped} since Re $|\omega _{EH}^{LF}|\gg $ Im $%
|\omega _{EH}^{LF}|$. Furthermore, in contrast with Ref.$\cite{1}$, the real
part Re $\omega _{EH}^{LF}$ is independent of $T$ and the damping rate Im $%
\omega _{EH}^{LF}$ is not quantized but varies as $T^{-3}$ or $T^{-2}$.

Due to specific properties of LFEH, such as, for instance, the essential
charge oscillations transversal to the edge, we may distinguish the LFEH
from the fundamental EMP of the $n=0$ LL. For {\it strong dissipation} the
latter mode is also called the HFEH of the $n=0$ LL. It is worth notice that
for the fundamental mode we have $2(Kk_{x}\ell _{0})^{2}\ll 1$ due to the
long wavelength condition $k_{x}\ell _{0}\ll 1$. Therefore $S_{j}$ can be
well approximated by its real part for the mode.

\subsection{Repulsion of fundamental EMPs for $\nu =4$}

Although here the condition $\Delta y_{01}/\ell _{0}\gg 1$ is well
justified, as can be seen in Fig. 1, and simplifies the treatment, the
system of Eqs. (\ref{8}) and (\ref{9}) can be strongly coupled by long-range
Coulomb interaction between the edges of the LL. If we neglect this
inter-edge Coulomb coupling in Eqs. (\ref{8}) and (\ref{9}), by setting the
coefficients $b_{mn}$ and $\tilde{b}_{mn}$ equal to zero, then Eq. (\ref{8})
leads to Eq. (\ref{15}) for the $\nu =2$ case and all edge modes of the $n=1$
LL are decoupled from those of the $n=0$ LL.

Assuming the latter, we first consider the symmetric edge modes of the $n=1$
LL. We consider both cases of {\it strong dissipation }and {\it weak
dissipation}. In the former case the fundamental mode of $n=1$ LL can be
called the HFEH of the $n=1$ LL. To treat the fundamental mode of $n=1$ LL
properly it is sufficient to consider the first two even terms in the sum
over $n$ in Eq. (\ref{7}), namely $\rho _{1}^{(0)}$ and $\rho _{1}^{(2)}$.
Then from Eq. (\ref{9}) for $m_{1}=0$ and $m_{1}=2$ we obtain a system of
two linear equations for $\rho _{1}^{(0)}$, $\rho _{1}^{(2)}$. The
corresponding dispersion relations of the HFEH of the $n=1$ LL becomes $%
\omega _{EH}^{(1)}\approx k_{x}v_{g1}+S_{1}(K-1/4)+S_{1}^{`}/12K$. The other
branch has the dispersion $\omega _{3}^{(1)}\approx
k_{x}v_{g1}+(S_{1}+2S_{1}^{`})/4$. In analogy with the fundamental EMP of
the $n=0$ LL, the $n=1$ LL fundamental EMP, decoupled from the excitations
of the $n=0$ LL, is very weakly damped even for {\it strong dissipation}.
Now if we omit the term of $\rho _{1}^{(2)}$ in Eq. (\ref{7}), i.e., if we
neglect the interaction between the monopole and quadrupole excitations of
the $n=1$ LL, the decoupled fundamental mode of $n=1$ LL has a dispersion
relation given by $\omega _{EH}^{(1)}$ without the damping term. This holds
for the $n=0$ LL as well, i.e., for its purely monopole excitation $\rho
_{0}^{(0)}$ with dispersion given by $\omega _{EH}^{(0)}$ without the
damping term.

If we take into account the Coulomb coupling between the pure monopole modes 
$\rho _{0}^{(0)}$ and $\rho _{1}^{(0)}$, their dispersions change
drastically. For $|k_{x}|\Delta y_{01}\ll 1$, the dispersion of the
renormalized fundamental mode of the $n=0$ LL becomes $\omega
_{+}^{(01)}\approx k_{x}(v_{g0}+v_{g1})/2+(2/\epsilon )k_{x}\tilde{\sigma}%
_{yx}^{0}[2\ln (1/k_{x}\ell _{0})-\ln (\Delta y_{01}/\ell _{0})+3/5]$ and
that of the $n=1$ LL $\omega _{-}^{(01)}\approx
k_{x}(v_{g0}+v_{g1})/2+(2/\epsilon )k_{x}\tilde{\sigma}_{yx}^{0}[\ln (\Delta
y_{01}/\ell _{0})+2/5]$. The dispersion relation $\omega _{+}^{(01)}(k_{x})$
is similar to that of the fundamental $j=0$ mode of Ref.$\cite{2}$ and to
the EMP of Ref.$\cite{1}$ since all of them have the term $\propto k_{x}\ln
(1/k_{x})$. Notice that $\omega _{+}^{(01)}(k_{x})$ is essentially different
from the frequency of the {\it decoupled} $n=0$ LL fundamental mode $\omega
\approx k_{x}v_{g0}+(2/\epsilon )\tilde{\sigma}_{yx}^{0}k_{x}[\ln
(1/k_{x}\ell _{0})+3/4]$. In contrast with $\omega _{+}^{(01)}(k_{x})$ the
dispersion of the renormalized fundamental EMP of the $n=1$ LL, $%
\omega_{-}^{(01)}(k_{x})$, becomes purely acoustic. Its phase velocity is
larger than that of the $j=1$ mode of Ref.$\cite{2}$ for $\Delta y_{01}/\ell
_{0}\geq 5$. Here we observe that the term related to the edge velocities, $%
k_{x}(v_{g0}+v_{g1})/2$, is typically much smaller than the second term
related to the electron-electron interaction. The spatial dependence of $%
\rho (\omega ,k_{x},y)$ for the renormalized fundamental mode of the $n=0$
LL is approximately $\propto [\Psi _{0}^{2}(\bar{y}_{0})+\Psi _{1}^{2}(\bar{y%
}_{1})]$ and that for the renormalized fundamental mode of the $n=1$ LL is $%
\propto [\Psi _{0}^{2}(\bar{y}_{0})-\Psi _{1}^{2}(\bar{y}_{1})]$.

The above results for Re $\omega (k_{x})$ of the coupled fundamental EMP of
the $n=0$ and $n=1$ LL's remain practically unchanged if more terms $%
\rho_{0}^{(j)}$ and $\rho _{1}^{(j)}$, for $j\geq 1$, are retained. Notice
that, due to the presence of the inter-edge Coulomb coupling, the symmetry
of the problem with respect to $y_{r0}$ or $y_{r1}$ is absent and hence, in
principle, $\rho _{0}^{(j)}$ and $\rho _{1}^{(j)}$ with $j$ even and odd,
must be included. Nevertheless, taking into account such additional terms
leads to essential contributions to the damping rates of the renormalized
fundamental EMPs of the $n=0$ and $n=1$ LLs. From $\Delta y_{01}/\ell_{0}\gg
1$, $k_{x}\Delta y_{01}\ll 1$ and the properties of the coefficients $a_{mn}$%
, $b_{mn}$, etc., we find that the most important terms in the sums of Eq. (%
\ref{7}) are $\rho _{0}^{(0)}$, $\rho _{0}^{(2)}$, $\rho _{1}^{(0)}$, and $%
\rho _{1}^{(2)}$. Calculations show that the {\it odd } terms here are
negligibly small. Dropping the less important term $\rho _{0}^{(2)}$, this
leads to a system of three coupled equations giving three branches $\tilde{%
\omega}_{\pm }^{(01)}$ and $\omega _{3}^{(01)}$. The dispersion relation of
the renormalized fundamental EMP of the $n=0$ LL is given by

\begin{equation}
\tilde{\omega}^{(01)}_{+}=k_x(v_{g0}+v_{g1})/2+ (2/\epsilon) k_{x} \tilde{%
\sigma}_{yx}^{0} [2 \ln(1/k_{x}\ell_{0})-\ln(\Delta y_{01}/\ell_{0})+3/5]+
S_{1}^{`}/16K  \label{17}
\end{equation}
and that of the $n=1$ LL by

\begin{equation}
\tilde{\omega}_{-}^{(01)}=k_{x}(v_{g0}+v_{g1})/2+(2/\epsilon )k_{x}\tilde{%
\sigma}_{yx}^{0}[\ln (\Delta y_{01}/\ell _{0})+2/5]+S_{1}^{`}/\{24[\ln
(\Delta y_{01}/\ell _{0})+\gamma +1/4]\}.  \label{18}
\end{equation}
The coupled fundamental EMPs $\omega _{\pm }^{(01)}$ are very weakly damped.
Notice that the neglect of the $\rho _{0}^{(2)}$ term in the calculation of
the damping rates is justified for substantially greater dissipation at the
edge of the $n=1$ LL than that at that of the $n=0$ LL. Indeed, $%
S_{j}^{`}\propto v_{gj}^{-4}$ and the group velocity $v_{g0}$ is typically
substantially larger than $v_{g1}$. So if we neglect the inter-edge Coulomb
coupling, the damping rate of the $\omega _{EH}^{(1)}$ branch is three times
larger than that of the $\omega _{EH}^{(0)}$ branch for $v_{g0}/v_{g1}=\sqrt{%
3}$.

\section{Discussion and Concluding Remarks}

We have introduced a realistic model for the confining potential $V_{y}$ and
took it sufficiently steep at the edge that LL flattening$^{\cite{9}}$ can
be discarded $^{\cite{10}\text{-}\cite{11}}$. For $\nu =2,4$, we have
neglected the spin splitting. This is a reasonable approximation in the bulk
of the channel but its validity near the edges is not clear in view of the
work of Refs.$\cite{15}$ and $\cite{16}$. We now compare our results
presented in Sec. III A with the experimental ones by Ashoori {\em et al}.$^{%
\cite{6}}$ for $\nu =1,\ T=0.3$K, and $B=5.1$ T. Using the experimental value%
$^{\cite{17}}$ $\Omega =7.8\times 10^{11}$/sec, we obtain $v_{g0}=8.8\times
10^{5}$ cm/sec. This leads to $\tilde{\sigma}_{yy}^{(0)}\propto T^{3}$. By
using a typical excited wave vector $q\simeq \pi /2L_{p}$, where $L_{p}=10\
\mu $m is the side of the square pulse, we find that all modes presented in
that section are very strongly damped except the fundamental mode $%
\omega_{EH}^{(0)}$ which is very weakly damped. Its damping rate is Im $%
\omega_{EH}^{(0)}\approx 2\times 10^{7}$/sec and its period of travel $%
T_{tr}\approx 3.4\times 10^{-9}$ sec, in agreement with the experimental
value. Here in the manner of Ref.$\cite{12}$ we have taken into account that
air is close to the 2DEG in the experiment$^{\cite{6}}$.

The dispersion relations given by Eqs. (\ref{17}) and (\ref{18}), for $\nu
=4 $, correspond to the experimental parameters $B=2.06$ T and $T=1.5$K of
Ref. $\cite{3}$. Using the experimentally determined$^{\cite{17}}$ $\Omega
=7.8\times 10^{11}$/sec gives $\Omega /\omega _{c}\approx 0.14$, $\Delta
y_{01}/\ell _{0}\approx 6$, $v_{g0}=2.3\times 10^{6}$/sec, and $%
v_{g0}/v_{g1}=\sqrt{3}$. The spectrum of the $\nu =4$ modes, shown in Fig.
(3a) of Ref. $\cite{3}$, is very well described by the dispersion of the
renormalized fundamental modes given by Eqs. (\ref{17}) and (\ref{18}).  The
same holds for the $\nu=4$ mode of Fig. 3 (b) of Ref. $\cite{3}$.  The mode $%
\omega_{3}^{(01)}$ is strongly damped.  Taking $\epsilon =6.75,$ its decay
rate is  Im $S_{1}^{`}/2\approx 2\tilde{\sigma}_{yy}^{(1)}/  \epsilon \ell
_{0}^{2}\approx 1.3\times 10^{10}$/sec.  The latter is still smaller than
that of the $j=1$  branch of Ref.$\cite{2}$, $1/\tau _{1}\approx 2\times
10^{10}$/sec,  which becomes four times larger for $B=1$T,  due to the $%
B^{-2}$ behavior. The decay rate of the $j=0$ mode of  Ref.$\cite{2}$ is $%
1/\tau _{0}\approx 1.7\times 10^{9}$/sec  whereas that of the $\tilde{\omega}%
_{+}^{(01)}$ mode,  given by Eq. (\ref{17}), is about ten times smaller,  Im 
$\tilde{\omega}_{+}^{(01)}\approx 2.1\times 10^{8}$/sec.  Moreover, the
damping rate of the $\tilde{\omega}_{-}^{(01)}$  mode, given by Eq. (\ref{18}%
), is much less than that of the $j=1$ mode of Ref. $\cite{2}$, since we
have  Im $\tilde{\omega}_{-}^{(01)}\approx 5.6\times 10^{8}$/sec $ \ll
1/\tau _{1}\approx 2\times 10^{10}$/sec.  Thus, the decay rates of the $%
\tilde{\omega}_{\pm }^{(01)}$  modes should be much closer to those of the 
experiment$^{\cite{3}}$ than the previous strongly  overestimated ones$^{%
\cite{2}}$. With regard to the delay  times $t_{d}$ for the sample with
length$^{\cite{3}}$ $L_{x}=320\mu $m, we obtain $t_{d}=1.2\times 10^{-10}$%
sec for the $\tilde{\omega}_{+}^{(01)}$ mode and $t_{d}=6.9\times 10^{-10}$%
sec  for the $\tilde{\omega}_{-}^{(01)}$ mode, in very good agreement  with
the experimental data$^{\cite{3}}$. From the previous  discussion, it
follows that the slower mode observed$^{\cite{3}}$  for $\nu =4$ is not the $%
j=1$ mode of Ref. $\cite{2}$ but  the present $\tilde{\omega}_{-}^{(01)}$
mode,  i.e., the renormalized fundamental EMP of the $n=1$ LL. It is also
clear  that our theory accounts for the existence of the  plateaus$^{\cite{3}%
}$ in $t_{d}$ as the quantized Hall conductivity  appears naturally in all
dispersion relations.

\acknowledgements

This work was supported by Brazilian FAPESP Grants No.98/10192-2 and
95/0789-3, Canadian NSERC Grant No. OGP0121756. In addition, O. G. B.
acknowledges partial support by the Ukrainian SFFI Grant No. 2.4/665, and N.
S. is grateful do Brazilian CNPq for a research fellowship.

$^{2}$E-mail: vbalev@power.ufscar.br

$^{3}$E-mail: takis@boltzmann.concordia.ca

$^{1}$E-mail: studart@power.ufscar.br

\begin{figure}[tbp]
\caption{Unperturbed electron density $n_{0}(y)$, normalized to the bulk
value $n_{0}$, as a function of $y/\ell _{0}$. The thick solid curve is the
model of Ref. 1 and the dotted curve that of Ref. 2 for $n_{0}(y)/n_{0}=(2/%
\pi )arctan[(y_{re}-y)/a]^{1/2}$, $a/\ell _{0}=20$. The dashed and solid
curves show the calculated profile for $\nu =1,2$ and for $\nu =4$,
respectively. The star and open dots denote the edges of the $n=1$ and $n=0$
LL's.}
\label{Fig. 1}
\end{figure}
\begin{figure}[tbp]
\caption{Dimensionless charge density profile $\tilde{\rho}(y) \equiv
\rho_{dip}$ for a dipole mode as a function of $\bar{y}_{0}/\ell _{0}$; $\nu
=2(1)$ and $\bar{n}=0$. The dashed, short-dashed, and solid curves
correspond to one ($l=1$), two ($l=1,3$), and three ($l=1,3,5$) terms,
respectively, kept in the expansion (\ref{7}). }
\label{Fig. 2}
\end{figure}
\begin{figure}[tbp]
\caption{Dimensionless charge density profile $\rho _{r}(y)=$Re$\delta \rho $
of the low-frequency edge helicon (LFEH) as a function of $\bar{y}_{0}/\ell
_{0}$ for $K/\eta =0.01$; $\nu =2(1)$, $\bar{n}=0$. The number of {\it even} 
$l$ terms retained in Eq. (\ref{7}) is shown next to the curves.}
\label{Fig. 3}
\end{figure}
\begin{figure}[tbp]
\caption{Dimensionless charge density profile $\rho _{i}(y)=$Im$\delta \rho $
of the low-frequency edge helicon (LFEH) as a function of $\bar{y}_{0}/\ell
_{0}$ for $K/\eta =0.01$; $\nu =2(1)$, $\bar{n}=0$. The number of {\it even} 
$l$ terms retained in Eq. (\ref{7}) is shown next to the curves.}
\label{Fig. 4}
\end{figure}
\begin{figure}[tbp]
\caption{Same as in Fig. 3 for $K/\eta =0.1$.}
\label{Fig. 5}
\end{figure}
\begin{figure}[tbp]
\caption{Same as in Fig. 4 for $K/\eta =0.1$.}
\label{Fig. 6}
\end{figure}

\end{document}